\documentclass{article}
\usepackage{spconf,amsmath,graphicx}
\usepackage{multirow}
\usepackage{array}
\usepackage{adjustbox}
\usepackage{booktabs}
\usepackage{amsthm,amsmath,amssymb}
\usepackage{mathrsfs}

\title{XWSB: A Blend System Utilizing XLS-R and WavLM with SLS Classifier detection system for SVDD 2024 Challenge}
\name{Qishan Zhang\textsuperscript{1}\textsuperscript{†}\thanks{†These authors contributed equally to this work.}, Shuangbing Wen\textsuperscript{2}\textsuperscript{†}, Fangke Yan\textsuperscript{1}\textsuperscript{†}, Tao Hu\textsuperscript{1,2}\textsuperscript{*}\thanks{*Corresponding author: Tao Hu(hutao\_es@hbmzu.edu.cn)},Member, IEEE, Jun Li\textsuperscript{1}}
\address{\textsuperscript{1}College of Intelligent Systems Science and Engineering, Hubei Minzu University, Enshi, China\\
\textsuperscript{2}School of Mathematics and Statistics, Hubei Minzu University, Enshi, China\\
}

\usepackage{url}
\begin{document}
%
\maketitle
\begin{abstract}


This paper introduces the model structure used in the SVDD 2024 Challenge.
The SVDD 2024 challenge has been introduced this year for the first time. 
Singing voice deepfake detection (SVDD) which faces complexities due to informal speech intonations and varying speech rates.
In this paper, we propose the XWSB system, which achieved SOTA performance in the SVDD challenge. XWSB stands for XLS-R, WavLM, and SLS Blend, representing the integration of these technologies for the purpose of SVDD. 
Specifically, we used the best performing model structure XLS-R\&SLS from the ASVspoof DF dataset, and applied SLS to WavLM to form the WavLM\&SLS structure. Finally, we integrated two models to form the XWSB system. Experimental results show that our system demonstrates advanced recognition capabilities in the SVDD challenge, specifically achieving an EER of 2.32\% in the CtrSVDD track. The code and data can be found at \url{https://github.com/QiShanZhang/XWSB_for_SVDD2024}.
\end{abstract}
\begin{figure*}[ht]
	
	\begin{minipage}[b]{1.0\linewidth}
		\centering
		\centerline{\includegraphics[width=17cm]{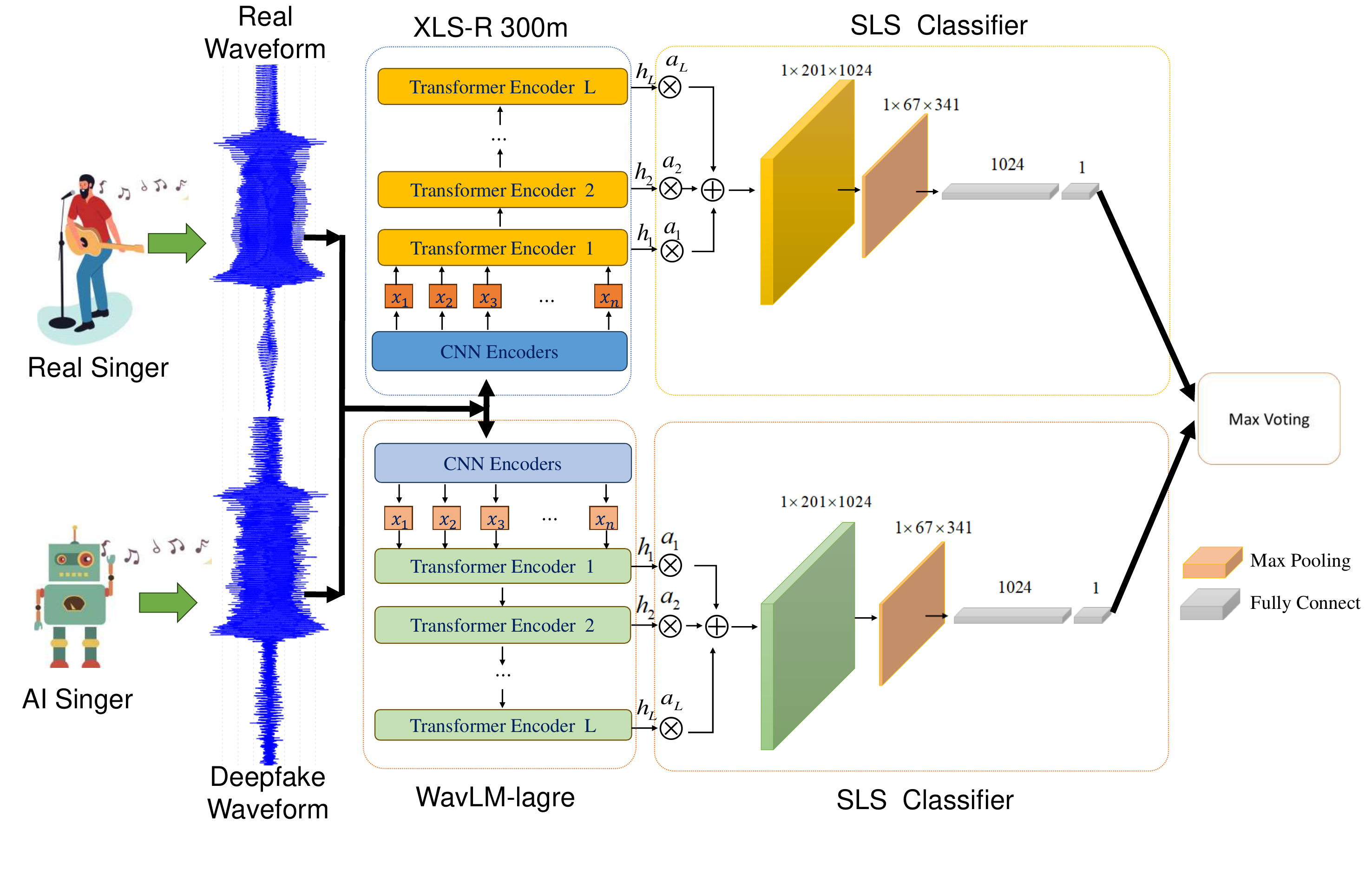}}
	\end{minipage}
	\caption{Pipeline of our Proposed model. Left part: the WavLM model and the XLS-R model; Right part: the SLS Classifier.}
	\label{fig1}
\end{figure*}

\section{Introduction}
\label{sec:intro}
The rise of virtual singers represents a significant advancement in the music sector, fueled by AI technology. Deep learning-based speech synthesis technology enables the synthesis of songs \cite{4}. Leveraging extensive training data and sophisticated neural network models, AI can generate highly realistic singing voices that capture the unique vocal characteristics and style of a singer. The primary technologies involved in AI singers are Singing Voice Synthesis (SVS) \cite{2} and Singing Voice Conversion (SVC) \cite{3, du2024dfadd}. While these technologies can enhance the auditory experience, they also present challenges, as AI singers infringe on the rights and interests of those whose voices are being imitated and raise concerns about identity and ethics \cite{1}. Thus, it is imperative to develop accurate voice forgery detection methods.

Since singing is also a type of human vocalization, we can leverage solutions from the field of deepfake speech detection \cite{5}. In recent years, the ASVspoof Challenge has become a pivotal event in the field of voice anti-spoofing, promoting the advancement and innovation of deepfake speech detection technologies \cite{30,31}. Deep spoofing detection for speech generally consists of a front-end and a back-end. The front-end extracts discriminative acoustic features from the audio, and the back-end employs classification models like Gaussian Mixture Models and Neural Networks to make a determination. Aakshi Mittal et al.\cite{6} proposed using static-dynamic CQCC features at the front-end and a hybrid deep learning model at the back-end for detecting spoofed audio. Haitao Yang et al. extracted two features, Mel spectrogram and Mel Frequency Cepstrum Coefficient (MFCC), at the front-end and utilized a two-stream network architecture with a feature fusion strategy at the back-end for speech detection. Nidhi Chakravarty et al.\cite{7} combined data enhancement techniques with a front-end hybrid feature extraction method, and the back-end employed Long Short-Term Memory (LSTM) and Gated Recurrent Unit (GRU) for detecting audio depth forgery.

However, current deep learning based methods still require special transformations of the audio. Therefore, self-supervised deep learning methods are introduced for detection. Self-supervised models for speech are able to extract a representation of the speech signal using a large amount of unlabelled data, thus enhancing the model's ability to detect forged speech \cite{8}. Jiang et al.\cite{9} proposed a self-supervised spoofed audio detection scheme that uses multiple convolutional blocks to extract audio signal features and a time-domain convolutional network (TCN) to extract contextual features, which confirms that self-supervised methods are effective for the forged audio detection task, but not as good as other DL methods in terms of performance. Pre-trained models based on self-supervised learning (SSL), such as wav2vec2.0 \cite{10}, HuBERT \cite{11}, WavLM \cite{12}, XLS-R \cite{16} and Whisper \cite{17}. These models have demonstrated excellent performance in various downstream tasks. Many researchers have utilized pre-trained models as front-end feature extractors for audio deep forgery detection. For example, Wang et al.\cite{13} first introduced self-supervised pre-trained speech models for deep audio forgery detection and obtained low EER scores on the ASVspoof 2019 LA test set. Li et al.\cite{14} proposed a self-supervised pre-trained model (HuBERT)-based approach for forgery speech detection on the ASVspoof2021 LA challenge database with an equal error rate (EER) of 2.89\%. Martín-Doñas et al. \cite{28} proposed a combination of a pre-trained wav2vec2 feature extractor and a downstream classifier to detect spoofed audio. Yinlin Guo et al. \cite{15} proposed the first audio deep forgery detection using a self-supervised WavLM and a multi-convergent attentional classifier, achieving EER 2.56\% at ASVspoof 2021 DF dataset. Zhang et al. \cite{zhang2024audio} proposed the XLS-R\&SLS model, where sensitive layer select (SLS) is a classifier. This model achieved an EER of 1.92 on the ASVspoof 2021 DF dataset and state-of-the-art performance on the In-The-Wild dataset. SingGraph \cite{chen24o_interspeech} is a novel model that also uses SSL models and achieves state-of-the-art performance on the SingFake dataset.

This paper is motivated by the following questions:
(i) Can models that perform well on the ASVspoof task continue to perform well on the SVDD task?
(ii) Can the SLS classifier be effectively applied to other pre-trained model architectures?
(iii) Do different pre-trained models exhibit varying performance, and can the fusion of multiple pre-trained models enhance overall performance?
By addressing these questions, we aim to explore the applicability and effectiveness of existing deepfake detection models and techniques in the new and challenging domain of singing voice deepfake detection.

The main contributions of this paper are as follows:
\begin{itemize}
\item[$\bullet$] We have applied the XLS-R\&SLS model architecture for the first time in the field of SVDD, demonstrating that models that perform well on the ASVspoof dataset can also achieve good performance after being trained in SVDD.
\item[$\bullet$] We first combined the SLS classifier with WavLM and achieved good performance on the CtrSVDD dataset, verifying the universality of the SLS module.
\item[$\bullet$] We proposed the XWSB (XLS-R-WavLM-SLS Blend) system, which achieves state-of-the-art EER in the CtrSVDD challenge track.
\end{itemize}

\section{METHOD}
\label{sec:format}
In this section, we provide a detailed description of the SLS classifier, the XLS-R and WavLM pre-trained models, and the modifications implemented to apply SLS to the SVDD task.

\subsection{Problem modeling}
We regard the SVDD task as a binary classification problem similar to the ASVspoof task; thus, we can employ the mathematical models used in the ASVspoof task for this purpose. To achieve this, let $x$ represent the waveform of an audio signal labeled $y \in \{0,1\}$, where $0$ indicates a synthetic human singing voice and $1$ indicates a real human singing voice. We aim to construct a classifier $\hat{y} = F_\theta(x)$ to predict the label of the input $x$. The binary detection model is constructed as a series of neural networks:
\begin{equation}
    F_\theta(x) = S_{\theta_S}(P_{\theta_P}(x))
\end{equation}
where $P_{\theta_P}$ represents the pre-train model for audio representation, equipped with its own set of parameters $\theta_P$. $S_{\theta_S}$ denotes a backend binary classifier with a SLS(Sensitive layer select) module designed to select useful outputs from different pre-train transformer layers, and $\theta_S$ are its parameters. This classifier can be optimized by solving:
\begin{equation}
    \min_\theta\sum_{(x,y) \in T}\mathcal{L}_b(y,F_\theta(x))
\end{equation}

where $\mathcal{L}_b(y,\hat{y})$ represents the Binary Focal Loss for binary classification, and $T$ denotes the training dataset comprising labeled real and synthetic examples. The model's overall framework is depicted in Fig. \ref{fig1}. The input raw waveform is fed into both the pre-trained WavLM model and the XLS-R model to obtain features from different layers. The rich hidden layer features from the two pre-trained models are then input into their respective SLS classifiers, which produce scores for the given singing audio. Finally, we employ a maximum voting method to integrate the two systems, selecting the score with the highest absolute value as the final detection result. This approach effectively highlights the most definitive predictions, thereby improving the accuracy of detecting forged singing voices.

\subsection{WavLM Model}

WavLM is a general speech pre-training model proposed by Microsoft Research Asia and the Microsoft Azure Speech Group \cite{9814838}, built on the HuBERT framework. The model consists of a convolutional encoder (CNN Encoder) and a Transformer encoder. The convolutional encoder has seven layers for preliminary processing and raw audio signal feature extraction. In the convolutional encoder, the original waveform is converted to a sequence of length N, $X = [x_ {1}, x_ {2},..., x_{N}]$, where N is the number of frames. In the Transformer encoder, gated relative position bias is used to incorporate relative position into the attention network calculation, thereby better modeling local information and improving the model's ability to capture short-term dependencies in speech signals. WavLM employs a masked speech denoising and prediction framework. Specifically, it creates a noisy environment by adding noise to the speech signal and randomly masking parts of the audio. Then, it uses self-supervised learning to predict pseudo-labels for the masked regions, thereby enhancing the model's ability to recognize and recover speech in noisy conditions. The Transformer encoder consists of multiple Transformer layers. The first layer receives features from the CNN encoder as input, and each subsequent layer processes the output from the previous layer. The output of each layer is represented as \({H}_i = [{h}_{1}, {h}_{2}, {h}_{3}, \cdots, {h}_{L}]\), where \({h}_L\) denotes the output at layer \(L\). 
WavLM is trained on 94,000 hours of unsupervised English data (including LibriLight, VoxPopuli, and GigaSpeech), during which it randomly transforms the input waveform. WavLM, combined with the MFA model, has demonstrated excellent performance on the ASVspoof 2021 dataset. Therefore, we chose to test the performance of WavLM combined with SLS.
\subsection{XLS-R Model}

XLS-R is a novel self-supervised model proposed by researchers at Facebook for various speech tasks \cite{16}. This model supports 128 languages and is trained on over 436,000 hours of public speech recordings (including VoxPopuli, Librispeech, CommonVoice, VoxLingua107, and BABEL). It is built on wav2vec 2.0 technology. XLS-R uses a multi-layer convolutional neural network to encode speech audio and employs a Transformer architecture to process the masked feature encoder's output, training through a contrastive learning task. The output of the Transformer layers is: $H = [h_{1}, h_{2},..., h_{L}]$. XLS-R can extract general and efficient speech features without a large amount of labeled data. The pre-trained XLS-R model can be fine-tuned for specific downstream tasks, achieving high performance with minimal labeled data. This gives XLS-R better generalization capabilities for multilingual tasks, while WavLM focuses more on high precision in single-language tasks and robustness to noise. Therefore, this paper uses both XLS-R and WavLM for front-end feature extraction to enhance the performance of singing voice deepfake detection tasks.
\subsection{SLS classifier}
\label{ssec:subhead}
The Sensitive Layer Select (SLS) classifier, proposed by Zhang et al.\cite{zhang2024audio}, is a classifier with an adaptive weight allocation module. It can adapt to transformer architectures with different numbers of layers, selecting appropriate weights between different layers of the transformer. This maximizes the utilization of all computational results from the transformer, accelerates convergence, and improves model accuracy. Its mathematical formulation is as follows:

\begin{equation}
F_\theta(x) = S_{\theta_S}(H) = \text{FC}(\text{maxpool}(\sum_{l=1}^{L} \alpha_l h_{l})) 
\end{equation}

where $h_{l}$ denotes the output of a distinct transformer layer in the pre-trained model, and $L$ represents the number of transformer layers. $\alpha_l$ represents the layer weight, defined as $\mathbf{\alpha} = [\alpha_1,\alpha_2,...,\alpha_{L}]$. This is derived from $H \in \mathbb{R}^{L \times N \times 1024}$, as illustrated in the equation below.

\begin{equation} 
\alpha = \text{Sigmoid}(\text{FC}(\text{avgpool}(H)))
\end{equation}

In the above equation, $avgpool$ denotes an average pooling operation that is applied along the dimension $N$ of $H \in \mathbb{R}^{L \times N \times 1024}$. Subsequently, ${H} \in \mathbb{R}^{L \times 1 \times 1024}$ is obtained, and through the fully connected layer (FC), ${H} \in \mathbb{R}^{L \times 1 \times 1}$ is produced.

\subsection{Model Ensemble}
Finally, we trained two separate models: one for deepfake audio detection using the pre-trained XLS-R model and SLS classifier, and another for deepfake audio detection using the pre-trained WavLM model and SLS classifier. Building on this, we employ an ensemble learning method using a max voting strategy. Specifically, we compare the absolute values of the scores output by the two models and choose the score with the larger absolute value as the final decision. This approach yields the final detection results.
\begin{equation}
S_{\text{final}} = \begin{cases}
S_{\text{X}} & \text{if } |S_{\text{X}}| = \max(|S_{\text{X}}|, |S_{\text{W}}|) \\
S_{\text{W}} & \text{if } |S_{\text{W}}| = \max(|S_{\text{X}}|, |S_{\text{W}}|)
\end{cases}
\end{equation}
where $S_{\text{final}}$ is the final score output by XWSB system, $S_{\text{X}}$ is the score output by XLS-R\&SLS model, $S_{\text{W}}$ is the score output by WavLM\&SLS model.

\begin{figure*}[htbp]
	
	\begin{minipage}[b]{1.0\linewidth}
		\centering
		\centerline{\includegraphics[width=18cm]{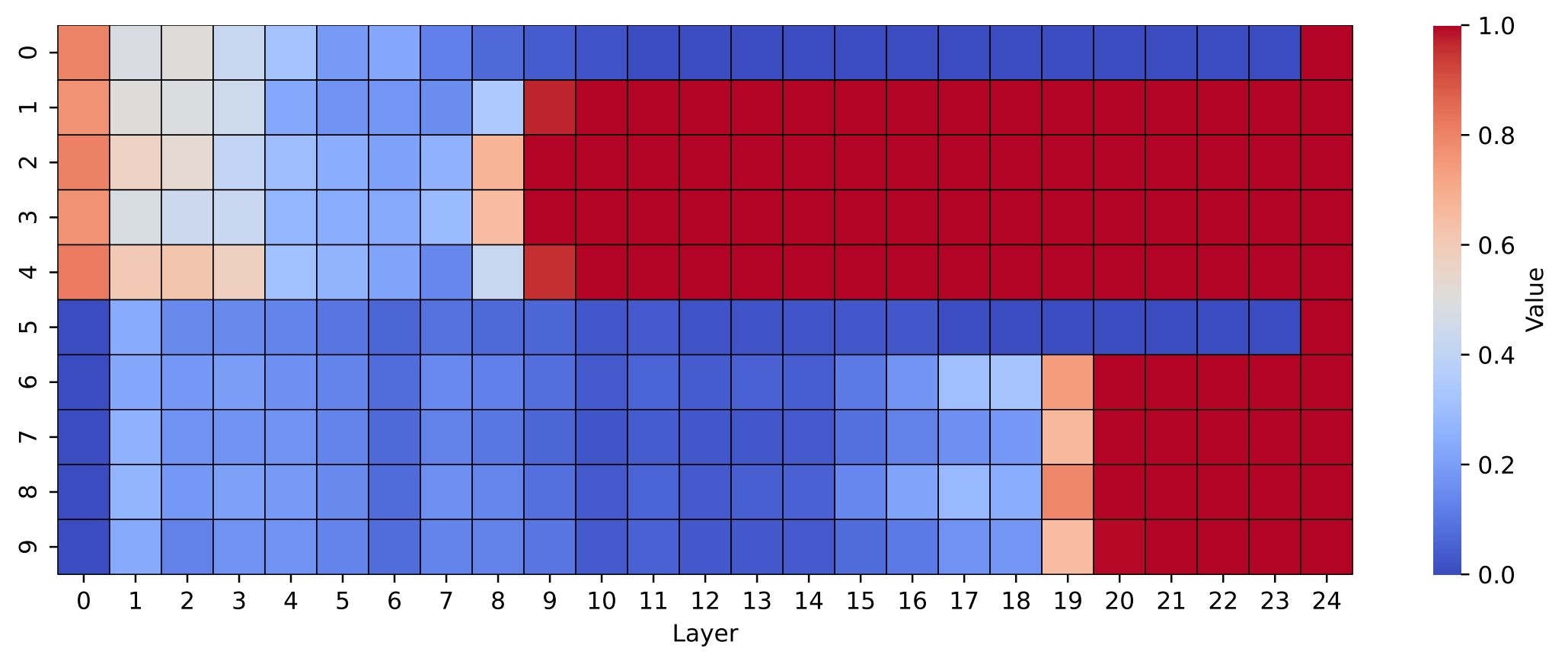}}
	\end{minipage}
	\caption{Randomly selected five test set audio samples, with the learned weights of the SLS module visualized. Rows 0-4 represent the combination of WavLM and SLS, while rows 5-9 represent the combination of XLS-R and SLS.}
	\label{fig2}
\end{figure*}
\section{EXPERIMENT}

In this section, we designed a series of experiments. First, we attempted to use the XLS-R\&SLS model, which has demonstrated state-of-the-art performance on the ASVspoof 2021 LA, DF, and In-the-Wild datasets \cite{19}. After training it on the competition training set, we submitted it for testing. Next, we replaced the pre-trained XLS-R model with WavLM. Finally, we combined the two separate models using the max voting method.

\subsection{Datasets and metrics}

The SVDD 2024 challenge consists of two tracks \cite{18}, and we describe the dataset for the CtrSVDD track. 
As shown in Table \ref{tab1}, we focus on the CtrSVDD track of the SVDD 2024 challenge.
The CtrSVDD dataset consists of 220,798 mono vocal clips in a total of 307.98 hours at a sample rate of 16 kHz. 
CtrSVDD includes 47.64 hours of genuine singing voices and 260.34 hours of deepfake singing voices, encompassing 14 deepfake methods and involving 164 singer identities. The test dataset includes voice segments from 48 speakers and 92,769 song clips. 
We use the Equal Error Rate (EER) to evaluate the performance of the SVDD system; the lower the EER, the better the system distinguishes between real and fake singing voices. In this experiment, we did not use any additional datasets, relying solely on the official training set, development set, and test set of the challenge.

\begin{table}[htbp]
\centering
\caption{The detailed information of the training and development partition of the CtrSVDD dataset.}
\label{tab1}   
\begin{adjustbox}{max width=\textwidth}
\scalebox{0.9}{
\begin{tabular}{lllll}
\hline
\multicolumn{1}{c}{\multirow{2}{*}{\textbf{Partition}}} & \multicolumn{1}{c}{\multirow{2}{*}{\textbf{Speakers}}} & \textbf{Bonafide}         & \multicolumn{1}{c}{\textbf{Deepfake}} & \multirow{2}{*}{\textbf{Attack Types}} \\ \cline{3-4}
\multicolumn{1}{c}{}                                    & \multicolumn{1}{c}{}                                   & \textbf{Utts}             & \multicolumn{1}{c}{\textbf{Utts}}     &                                        \\ \hline
Train                                                   & 59                                                     & \multicolumn{1}{l}{12169} & 72235                                 & A01$\sim$A08                           \\
Dev                                                     & 55                                                     & \multicolumn{1}{l}{6547}  & 37078                                 & A01$\sim$A08                           \\
Test                                                    & 48                                                 & 13,596&79,173& A01$\sim$A14                                                \\ \hline
\end{tabular}}
\end{adjustbox}
\end{table}

\begin{table*}[ht]
\label{1}
 \caption{Final results in terms of EER (\%) of our submitted approaches to the  CtrSVDD  2024 challenge. lncludes different variants for the XLS-R \& SLS and the WavLM \& SLS and XWSB.}
  \centering
 \scalebox{0.9}{
  \begin{tabular}{lccccccccccc}
    \toprule          
Model    &A09&A10&A11&A12&A13&A14&Result (overall)&Result (w/o ACESinger)\\
    \midrule
B01 &-  & -& -& -& -& -& 16.10& 12.03\\
B02&-  & -& -& -& -& -&  13.58& 11.16\\
XLS-R \& SLS &1.76  & 1.19& 4.33& 6.35& 1.22& 11.94& 5.12& 3.73\\
WavLM \& SLS & 1.39  & 0.87& 2.66& 5.09& 0.92& 11.59&4.92 & 2.59 \\
    \textbf{XWSB} &\textbf{1.02}&\textbf{0.69}& \textbf{2.54}& \textbf{4.42}& \textbf{0.76}&\textbf{11.35}& \textbf{4.45}& \textbf{2.32}& \\
    \bottomrule
   \end{tabular}}
  \label{tab2}
\end{table*}

\subsection{Experiment Setup}
\label{sec:}

Our models were implemented using PyTorch and trained on a single GeForce RTX 4090 GPU. For audios exceeding four seconds, we randomly selected a 4-second segment from the clip. For audios shorter than four seconds, we repeated the audio to reach 4 seconds. These 4-second raw waveforms were fed into the pre-trained WavLM model and XLS-R model. These models were jointly trained with the SLS classifier using the AdamW optimizer. The learning rate for the WavLM model was set to \(1 \times 10^{-5}\), and the learning rate for the XLS-R model was set to \(5 \times 10^{-7}\), with a weight decay coefficient of \(1 \times 10^{-9}\). Additionally, we employed a cosine annealing learning rate scheduler with a maximum T-value of 10 and a minimum learning rate of \(1 \times 10^{-6}\). The batch size was set to 5. The WavLM model used was WavLM Large, and the XLS-R model used was XLS-R(0.3B).

\subsection{Experiment result for CtrSVDDtrack and analysis}
Table\ref{tab2} presents the experimental results of the systems we used for the SVDD task, compared with the official baseline. Both models we trained achieved state-of-the-art (SOTA) performance in the CtrSVDD track of the challenge. Moreover, the model integration scheme we employed further improved the model's performance.

As shown in Table \ref{tab2}, both XLS-R and WavLM, when combined with the SLS module, demonstrate cutting-edge recognition accuracy. However, WavLM performed better than XLS-R across several attacks, and we believe this is due to insufficient training of XLS-R. When the two models are integrated using the majority voting method, the resulting accuracy is not simply the highest accuracy of the two models, but rather an improvement over the original individual systems.
\begin{table}[htbp]
\label{DFLAresult}
 \caption{Comparative Pooled EER(\%) results of our proposed method with other systems in the dataset aspect.}
  \centering
  \scalebox{1}{
  \begin{tabular}{lcccc}
    \toprule          
Model    &kising&m4singer&acesinger\\
    \midrule
XLS-R \& SLS &3.96  & 3.73& 50.58\\
WavLM \& SLS & 5.44  & 2.59& 50.40&  \\
    \textbf{XWSB} &\textbf{2.82}&\textbf{2.32}& \textbf{50.05} \\
    \bottomrule
  \end{tabular}}
  \label{tab3}
\end{table}

As shown in Table \ref{tab3}, the performance of the two models on different datasets varies significantly. Similarly, the integration of the models demonstrates superior results compared to the individual models. However, when faced with the Acesinger dataset, neither model could distinguish between genuine and fake singing voices, and the integration did not improve performance. This is an issue that warrants further analysis in the future.

Figure \ref{fig2} shows the inference results of randomly selecting five audio samples from the test set using two models. Although the SLS module was trained separately with WavLM and XLS-R, it unexpectedly exhibited the same pattern when processing the same audio (Row 0 and Row 5).While the other four rows differ in the number of layer selections, they all use similar patterns.
According to the labels publicly provided by \cite{zhang2024svdd}, it is confirmed that the audio in Row 1 is bona-fide, while the others are deepfake.
 Another possible reason why WavLM outperforms XLS-R is that WavLM outputs convolutional layer features. Effectively utilizing hidden layer features remains a significant area for research.
\label{sec:typestyle}

\section{Conclusion}
\label{sec:print}
In this paper, we validated the hypothesis that models performing well on the ASVspoof dataset can still exhibit good performance after being trained with SVDD data. We tested the model, which achieved state-of-the-art performance on the ASVspoof dataset, in the SVDD challenge. After training with the SVDD challenge training set without adding any additional datasets, the model still demonstrated excellent performance. This further proves the general adaptability of the SLS classifier combined with other pre-trained models. We proposed the XWSB system, which exhibited SOTA performance in the CtrSVDD track of the SVDD challenge. Future work will focus on the In-The-Wild track and aim to improve the simple structure of the SLS classifier to further enhance the model's performance. Further ablation experiments may be necessary.
\section{ACKNOWLEDGEMENT}
This work was supported by the Outstanding Youth Science and Technology Innovation Team Project for Colleges and Universities of Hubei Province of China
(T2023013) and Natural Science Foundation of Hubei Province of China under Grant 2023AFD061.

\bibliographystyle{IEEEbib}
\bibliography{output}

\end{document}